\documentclass[11pt,a4paper]{article}
\usepackage{acl2015}
\usepackage{times}
\usepackage{latexsym}
\usepackage{amsmath}
\usepackage{multirow}
\usepackage{url}
\usepackage{tabularx}
\usepackage{ragged2e}  
\usepackage{graphicx}
\usepackage{caption}
\usepackage{subcaption}

\title{Enhanced Twitter Sentiment Classification Using Contextual Information}

\author{Soroush Vosoughi \\
The Media Lab\\ MIT\\
Cambridge, MA 02139\\
{\small \tt soroush@mit.edu}\\
\And Helen Zhou \\
The Media Lab\\ MIT\\
Cambridge, MA 02139\\
{\small \tt hlzhou@mit.edu}\\
\And Deb Roy \\
The Media Lab\\ MIT\\
Cambridge, MA 02139\\
{\small \tt dkroy@media.mit.edu}\\
}

\date{}

\begin{document}
\maketitle

\begin{abstract}
The rise in popularity and ubiquity of Twitter has made sentiment analysis of tweets an important and well-covered area of research. However, the 140 character limit imposed on tweets makes it hard to use standard linguistic methods for sentiment classification. On the other hand, what tweets lack in structure they make up with sheer volume and rich metadata. This metadata includes geolocation, temporal and author information. We hypothesize that sentiment is dependent on all these contextual factors. Different locations, times and authors have different emotional valences. In this paper, we explored this hypothesis by utilizing distant supervision to collect millions of labelled tweets from different locations, times and authors. We used this data to analyse the variation of tweet sentiments across different authors, times and locations. Once we explored and understood the relationship between these variables and sentiment, we used a Bayesian approach to combine these variables with more standard linguistic features such as n-grams to create a Twitter sentiment classifier. This combined classifier outperforms the purely linguistic classifier, showing that integrating the rich contextual information available on Twitter into sentiment classification is a promising direction of research.

\end{abstract}



\section{Introduction}
Twitter is a micro-blogging platform and a social network where users can publish and exchange short messages of up to 140 characters long (also known as tweets). Twitter has seen a great rise in popularity in recent years because of its availability and ease-of-use. This rise in popularity and the public nature of Twitter (less than 10\% of Twitter accounts are private \cite{techcrunch2009}) have made it an important tool for studying the behaviour and attitude of people.

One area of research that has attracted great attention in the last few years is that of tweet sentiment classification. Through sentiment classification and analysis, one can get a picture of people's attitudes about particular topics on Twitter. This can be used for measuring people's attitudes towards brands, political candidates, and social issues. 

There have been several works that do sentiment classification on Twitter using standard sentiment classification techniques, with variations of n-gram and bag of words being the most common. There have been attempts at using more advanced syntactic features as is done in sentiment classification for other domains \cite{read2005using,nakagawa2010dependency}, however the 140 character limit imposed on tweets makes this hard to do as each article in the Twitter training set consists of sentences of no more than several words, many of them with irregular form \cite{saif2012alleviating}.

On the other hand, what tweets lack in structure they make up with sheer volume and rich metadata. This metadata includes geolocation, temporal and author information. We hypothesize that sentiment is dependent on all these contextual factors. Different locations, times and authors have different emotional valences. For instance, people are generally happier on weekends and certain hours of the day, more depressed at the end of summer holidays, and happier in certain states in the United States. Moreover, people have different baseline emotional valences from one another. These claims are supported for example by the annual Gallup poll that ranks states from most happy to least happy \cite{gallup2014}, or the work by Csikszentmihalyi and Hunter \cite{csikszentmihalyi2003happiness} that showed reported happiness varies significantly by day of week and time of day. We believe these factors manifest themselves in sentiments expressed in tweets and that by accounting for these factors, we can improve sentiment classification on Twitter. 

In this work, we explored this hypothesis by utilizing \emph{distant supervision} \cite{go2009twitter} to collect millions of labelled tweets from different locations (within the USA), times of day, days of the week, months and authors. We used this data to analyse the variation of tweet sentiments across the aforementioned categories. We then used a Bayesian approach to incorporate the relationship between these factors and tweet sentiments into standard n-gram based Twitter sentiment classification.

This paper is structured as follows. In the next sections we will review related work on sentiment classification, followed by a detailed explanation of our approach and our data collection, annotation and processing efforts. After that, we describe our baseline n-gram sentiment classifier model, followed by the explanation of how the baseline model is extended to incorporate contextual information. Next, we describe our analysis of the variation of sentiment within each of the contextual categories. We then evaluate our models and finally summarize our findings and contributions and discuss possible paths for future work.





\section{Related Work}
Sentiment analysis and classification of text is a problem that has been well studied across many different domains, such as blogs, movie reviews, and product reviews (e.g., \cite{pang2002thumbs,cui2006comparative,chesley2006using}). There is also extensive work on sentiment analysis for Twitter. Most of the work on Twitter sentiment classification either focuses on different machine learning techniques (e.g., \cite{wang2011topic,jiang2011target}), novel features (e.g., \cite{davidov2010enhanced,kouloumpis2011twitter,saif2012alleviating}), new data collection and labelling techniques (e.g., \cite{go2009twitter}) or the application of sentiment classification to analyse the attitude of people about certain topics on Twitter (e.g., \cite{diakopoulos2010characterizing,bollen2011modeling}). These are just some examples of the extensive research already done on Twitter sentiment classification and analysis.

There has also been previous work on measuring the happiness of people in different contexts (location, time, etc). This has been done mostly through traditional land-line polling \cite{csikszentmihalyi2003happiness,gallup2014}, with Gallup's annual happiness index being a prime example \cite{gallup2014}. More recently, some have utilized Twitter to measure people's mood and happiness and have found Twitter to be a generally good measure of the public's overall happiness, well-being and mood. For example, Bollen et al. \cite{bollen2011modeling} used Twitter to measure the daily mood of the public and compare that to the record of social, political, cultural and economic events in the real world. They found that these events have a significant effect on the public mood as measured through Twitter. Another example would be the work of Mitchell et al. \cite{mitchell2013geography}, in which they estimated the happiness levels of different states and cities in the USA using Twitter and found statistically significant correlations between happiness level and the demographic characteristics (such as obesity rates and education levels) of those regions. Finally, improving natural language processing by incorporating contextual information has been successfully attempted before \cite{vosoughi2014improving,roy2014grounding}; but as far as we are aware, this has not been attempted for sentiment classification.

In this work, we combined the sentiment analysis of different authors, locations, times and dates as measured through labelled Twitter data with standard word-based sentiment classification methods to create a context-dependent sentiment classifier. As far as we can tell, there has not been significant previous work on Twitter sentiment classification that has achieved this.



\section{Approach}
The main hypothesis behind this work is that the average sentiment of messages on Twitter is different in different contexts. Specifically, tweets in different spatial, temporal and authorial contexts have on average different sentiments. Basically, these factors (many of which are environmental) have an affect on the emotional states of people which in turn have an effect on the sentiments people express on Twitter and elsewhere. In this paper, we used this contextual information to better predict the sentiment of tweets. 

Luckily, tweets are tagged with very rich metadata, including location, timestamp, and author information. By analysing labelled data collected from these different contexts, we calculated \emph{prior} probabilities of negative and positive sentiments for each of the contextual categories shown below:

\begin{itemize}
\item The states in the USA (50 total).
\item Hour of the day (HoD) (24 total).
\item Day of week (DoW) (7 total).
\item Month (12 total).
\item Authors (57710 total).
\end{itemize}

This means that for every item in each of these categories, we calculated a probability of sentiment being positive or negative based on historical tweets. For example, if seven out of ten historical tweets made on Friday were positive then the prior probability of a sentiment being positive for tweets sent out on Friday is $0.7$ and the prior probability of a sentiment being negative is $0.3$. We then trained a Bayesian sentiment classifier using a combination of these prior probabilities and standard n-gram models. The model is described in great detail in the "Baseline Model" and "Contextual Model" sections of this paper.

In order to do a comprehensive analysis of sentiment of tweets across aforementioned contextual categories, a large amount of labelled data was required. We needed thousands of tweets for every item in each of the categories (e.g. thousands of tweets per hour of day, or state in the US). Therefore, creating a corpus using human-annotated data would have been impractical. Instead, we turned to distant supervision techniques to obtain our corpus. Distant supervision allows us to have noisy but large amounts of annotated tweets. 

There are different methods of obtaining labelled data using distant supervision \cite{read2005using,go2009twitter,barbosa2010robust,davidov2010enhanced}. We used emoticons to label tweets as positive or negative, an approach that was introduced by Read \cite{read2005using} and used in multiple works \cite{go2009twitter,davidov2010enhanced}. We collected millions of English-language tweets from different times, dates, authors and US states. We used a total of six emoticons, three mapping to positive and three mapping to negative sentiment (table \ref{tab:emot}). We identified more than 120 positive and negative ASCII emoticons and unicode emojis\footnote{Japanese pictographs similar to ASCII emoticons}, but we decided to only use the six most common emoticons in order to avoid possible selection biases. For example, people who use obscure emoticons and emojis might have a different base sentiment from those who do not. Using the six most commonly used emoticons limits this bias. Since there are no "neutral" emoticons, our dataset is limited to tweets with positive or negative sentiments. Accordingly, in this work we are only concerned with analysing and classifying the polarity of tweets (negative vs. positive) and not their subjectivity (neutral vs. non-neutral). Below we will explain our data collection and corpus in greater detail.

\begin{table} [htbp]
\centering
\begin{tabular}{|c|c|}
\hline
Positive Emoticons  & Negative Emoticons  \\ \hline
:)  & :( \\ 
:-)  & :-( \\ 
: ) & : ( \\
\hline

\end{tabular}
\caption{List of emoticons.}
\label{tab:emot}
\end{table}

\section{Data Collection and Datasets}
We collected two datasets, one massive and labelled through distant supervision, the other small and labelled by humans. The massive dataset was used to calculate the prior probabilities for each of our contextual categories. Both datasets were used to train and test our sentiment classifier. The human-labelled dataset was used as a sanity check to make sure the dataset labelled using the emoticons classifier was not too noisy and that the human and emoticon labels matched for a majority of tweets.

\subsection{Emoticon-based Labelled Dataset}
We collected a total of $18$ million, geo-tagged, English-language tweets over three years, from January 1st, 2012 to January 1st, 2015, evenly divided across all 36 months, using Historical PowerTrack for Twitter\footnote{Historical PowerTrack for Twitter provides complete access to the full archive of Twitter public data.} provided by GNIP\footnote{https://gnip.com/}. We created geolocation bounding boxes\footnote{The bounding boxes were created using http://boundingbox.klokantech.com/} for each of the 50 states which were used to collect our dataset. All $18$ million tweets originated from one of the 50 states and are tagged as such. Moreover, all tweets contained one of the six emoticons in Table \ref{tab:emot} and were labelled as either positive or negative based on the emoticon. Out of the $18$ million tweets, $11.2$ million ($62\%$) were labelled as positive and $6.8$ million ($38\%$) were labelled as negative. The $18$ million tweets came from $7,657,158$ distinct users.

\subsection{Human Labelled Dataset}
We randomly selected $3000$ tweets from our large dataset and had all their emoticons stripped. We then had these tweets labelled as positive or negative by three human annotators. We measured the inter-annotator agreement using \emph{Fleiss' kappa}, which calculates the degree of agreement in classification over that which would be expected by chance \cite{fleiss1971measuring}. The \emph{kappa} score for the three annotators was $0.82$, which means that there were disagreements in sentiment for a small portion of the tweets. However, the number of tweets that were labelled the same by at least two of the three human annotator was $2908$ out of of the $3000$ tweets ($96\%$). Of these $2908$ tweets, $60\%$ were labelled as positive and $40\%$ as negative.

We then measured the agreement between the human labels and emoticon-based labels, using only tweets that were labelled the same by at least two of the three human annotators (the majority label was used as the label for the tweet). Table \ref{tab:conf} shows the confusion matrix between human and emoticon-based annotations. As you can see, $85\%$ of all labels matched ($\frac{1597+822}{1597+882+281+148}=.85$).

\begin{table}[htbp]
\centering
\begin{tabular}{l|c|c}
 & Human-Pos & Human-Neg  \\ \hline
Emot-Pos & $1597$  & $281$  \\ \hline
Emot-Neg & $148$ &  $882$
\end{tabular}
\caption{Confusion matrix between human-labelled and emoticon-labelled tweets.}
\label{tab:conf}
\end{table}

These results are very promising and show that using emoticon-based distant supervision to label the sentiment of tweets is an acceptable method. Though there is some noise introduced to the dataset (as evidenced by the $15\%$ of tweets whose human labels did not match their emoticon labels), the sheer volume of labelled data that this method makes accessible, far outweighs the relatively small amount of noise introduced.

\subsection{Data Preparation}
Since the data is labelled using emoticons, we stripped all emoticons from the training data. This ensures that emoticons are not used as a feature in our sentiment classifier. A large portion of tweets contain links to other websites. These links are mostly not meaningful semantically and thus can not help in sentiment classification. Therefore, all links in tweets were replaced with the token  "URL". Similarly, all mentions of usernames (which are denoted by the \emph{@} symbol) were replaced with the token "USERNAME", since they also can not help in sentiment classification. Tweets also contain very informal language and as such, characters in words are often repeated for emphasis (e.g., the word \emph{good} is used with an arbitrary number of \emph{o}'s in many tweets). Any character that was repeated more than two times was removed (e.g., \emph{goooood} was replaced with \emph{good}). Finally, all words in the tweets were stemmed using \emph{Porter Stemming} \cite{porter1980algorithm}.


\section{Baseline Model}
For our baseline sentiment classification model, we used our massive dataset to train a negative and positive n-gram language model from the negative and positive tweets. 

%


As our baseline model, we built purely linguistic bigram models in Python, utilizing some components from NLTK \cite{bird2009natural}. These models used a vocabulary that was filtered to remove words occurring 5 or fewer times. Probability distributions were calculated using Kneser-Ney smoothing \cite{chen1999empirical}. 
In addition to Kneser-Ney smoothing, the bigram models also used ``backoff'' smoothing \cite{katz1987estimation}, in which an n-gram model falls back on an $(n-1)$-gram model for words that were unobserved in the n-gram context.

In order to classify the sentiment of a new tweet, its probability of fit is calculated using both the negative and positive bigram models. 
Equation \ref{eq-1} below shows our models through a Bayesian lens.

\begin{align}
  &\begin{aligned}  
\Pr(\theta_{s}\mid W) &=
 &\frac{\Pr(W\mid\theta_{s}) \Pr(\theta_{s})}{\Pr(W)}
  \end{aligned}
  \label{eq-1}
\end{align}

Here $\theta_{s}$ can be $\theta_{p}$ or $\theta_{n}$, corresponding to the hypothesis that the sentiment of the tweet is positive or negative respectively. $W$ is the sequence of $\ell$ words, written as $w_1^\ell$, that make up the tweet. $\Pr(W)$ is not dependent on the hypothesis, and can thus be ignored. Since we are using a bigram model, Equation \ref{eq-1} can be written as:


\begin{align}
  &\begin{aligned}
\Pr&(\theta_{s}\mid W) \propto
 &\prod_{i=2}^\ell \Pr(w_i\mid w_{i-1},\theta_{s}) \Pr(\theta_{s})
\end{aligned}
\label{eq-2}
\end{align}



This is our purely linguistic baseline model.

\section{Contextual Model}
The Bayesian approach allows us to easily integrate the contextual information into our models. $\Pr(\theta_{s})$ in Equation \ref{eq-2} is the prior probability of a tweet having the sentiment $s$. The prior probability ($\Pr(\theta_{s})$) can be calculated using the contextual information of the tweets. Therefore, $\Pr(\theta_{s})$ in equation \ref{eq-2} is replaced by $\Pr(\theta_{s}|C)$, which is the probability of the hypothesis given the contextual information. $\Pr(\theta_{s}|C)$ is the posterior probability of the following Bayesian equation:

\begin{equation} 
\Pr(\theta_{s}\mid C) = \frac{\Pr(C\mid \theta_{s}) \Pr(\theta_{s})}{\Pr(C)}
\label{eq-4}
\end{equation}
Where $C$ is the set of contextual variables: $\{State, HoD, Dow, Month, Author\}$. $\Pr(\theta_{s}|C)$ captures the probability that a tweet is positive or negative, given the state, hour of day, day of the week, month and author of the tweet. Here $\Pr(C)$ is not dependent on the hypothesis, and thus can be ignored. Equation \ref{eq-2} can therefore be rewritten to include the contextual information:

\begin{align}
  &\begin{aligned}
\Pr(\theta_{s}\mid W,C) \propto& \prod_{i=2}^\ell \Pr(w_i\mid w_{i-1},\theta_{s}) \\
 &\Pr(C\mid \theta_{s}) \Pr(\theta_{s})
\end{aligned}
\label{eq-5}
\end{align}



Equation \ref{eq-5} is our extended Bayesian model for integrating contextual information with more standard, word-based sentiment classification.

\section{Sentiment in Context}
We considered five contextual categories: one spatial, three temporal and one authorial. Here is the list of the five categories:

\begin{itemize}
\item The states in the USA (50 total) (spatial).
\item Hour of the day (HoD) (24 total) (temporal).
\item Day of week (DoW) (7 total) (temporal).
\item Month (12 total) (temporal).
\item Authors (57,710 total) (authorial).
\end{itemize}

We used our massive emoticon labelled dataset to calculate the average sentiment for all of these five categories. A tweet was given a score of $-1$ if it was labelled as negative and a score $1$ if it was labelled as positive, so an average sentiment of $0$ for a contextual category would mean that tweets in that category were evenly labelled as positive and negative.

\subsection{Spatial}

\begin{figure}[htbp]
\centering
\includegraphics[width=1.0\columnwidth]{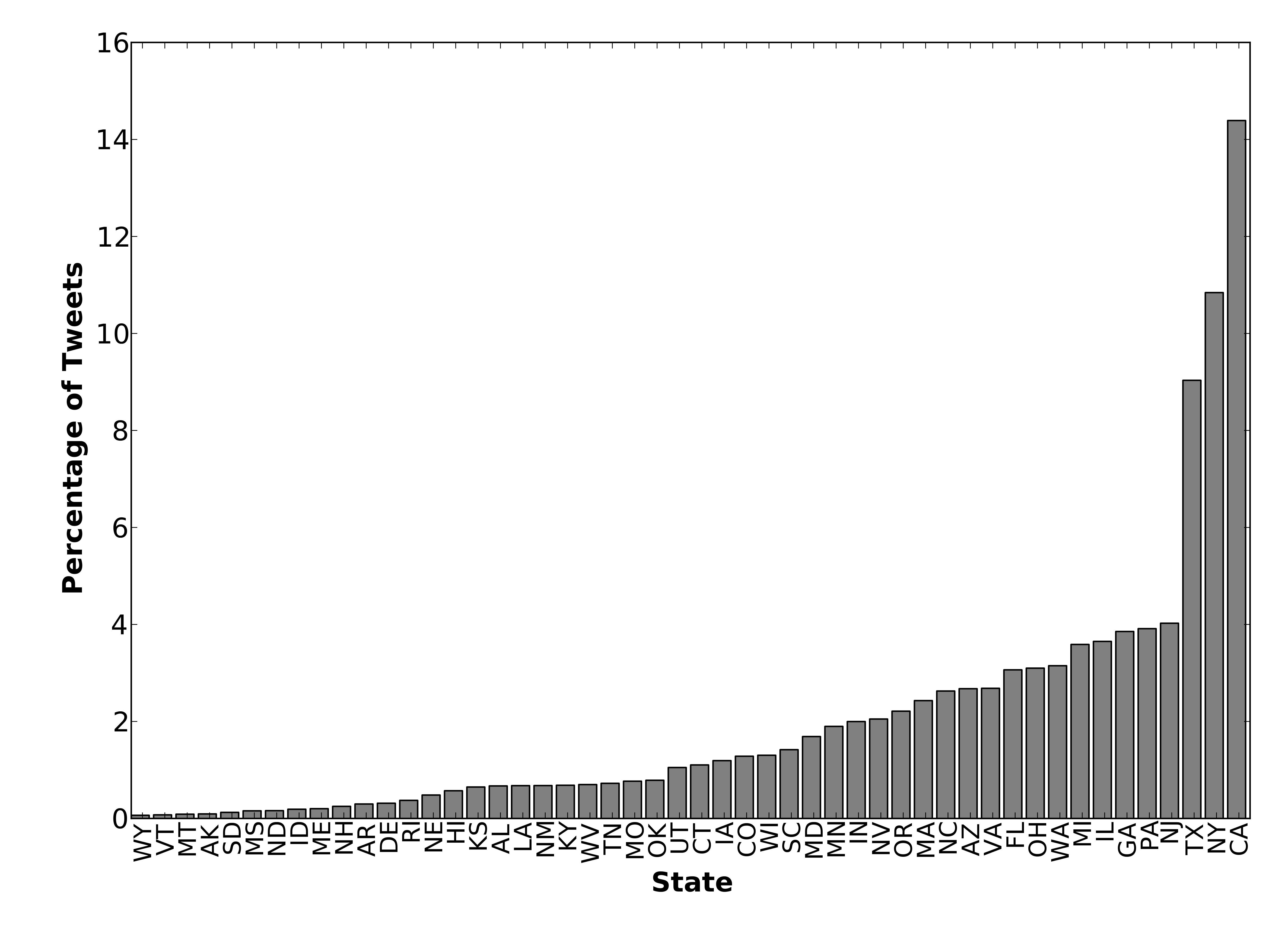}
\caption{Percentage of tweets per state in the USA, sorted from lowest to highest.}
\label{fig:scount}
\end{figure}

All of the $18$ million tweets in our dataset originate from the USA and are geo-tagged. Naturally, the tweets are not evenly distributed across the 50 states given the large variation between the population of each state. Figure \ref{fig:scount} shows the percentage of tweets per state, sorted from smallest to largest. Not surprisingly, California has the highest number of tweets ($2,590,179$), and Wyoming has the lowest number of tweets ($11,719$). 

\begin{figure}[h]
\centering
\includegraphics[width=1.0\columnwidth]{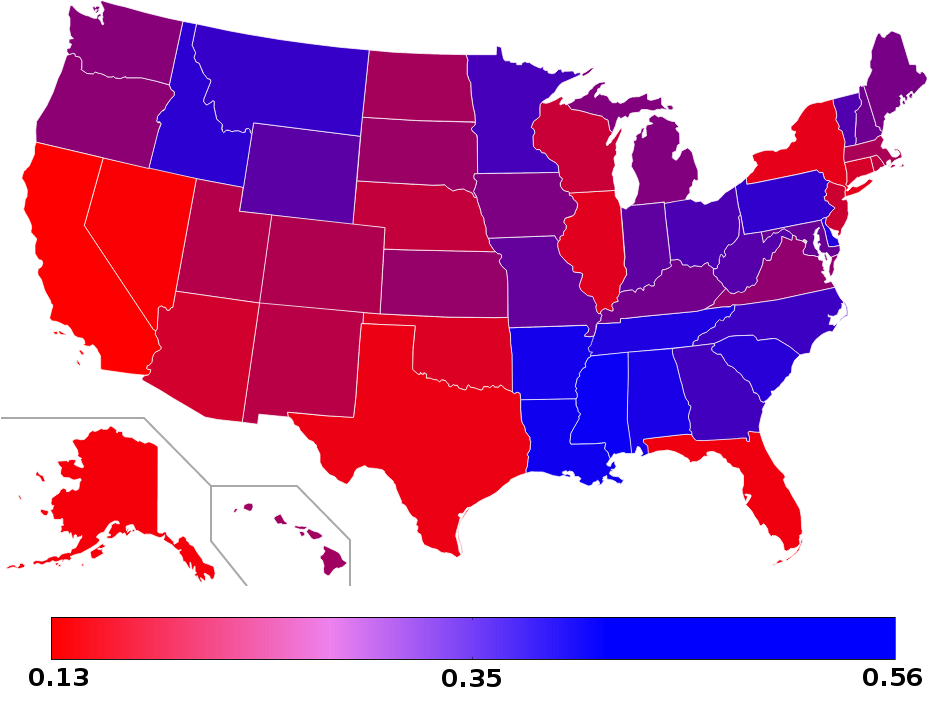}
\caption{Average sentiment of states in the USA, averaged across three years, from 2012 to 2014.}
\label{fig:map}
\end{figure}

Even the state with the lowest percentage of tweets has more than ten thousand tweets, which is enough to calculate a statistically significant average sentiment for that state. The sentiment for all states averaged across the tweets from the three years is shown in Figure \ref{fig:map}. Note that an average sentiment of $1.0$ means that all tweets were labelled as positive, $-1.0$ means that all tweets were labelled as negative and $0.0$ means that there was an even distribution of positive and negative tweets. The average sentiment of all the states leans more towards the positive side. This is expected given that $62\%$ of the tweets in our dataset were labelled as positive.

\begin{figure}[h]
\centering
\includegraphics[width=1.0\columnwidth]{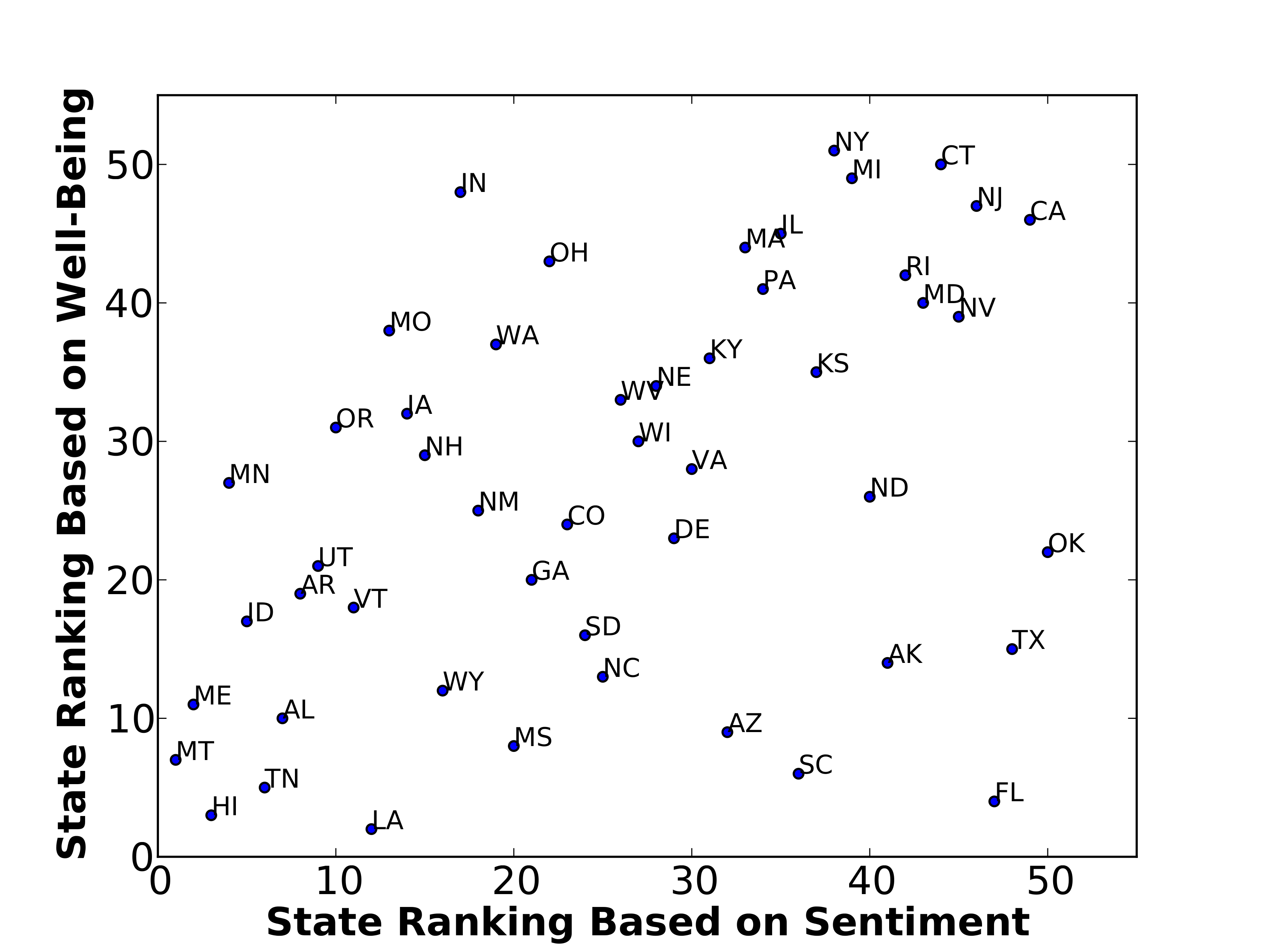}
\caption{Ranking of US states based on Twitter sentiment vs. ranking of states based on their well-being index. $r=0.44$, $p<0.005$. }
\label{fig:ranking}
\end{figure}

It is interesting to note that even with the noisy dataset, our ranking of US states based on their Twitter sentiment correlates with the ranking of US states based on the well-being index calculated by Oswald and Wu \cite{oswald2011well} in their work on measuring well-being and life satisfaction across America. Their data is from the behavioral risk factor survey score (BRFSS), which is a survey of life satisfaction across the United States from $1.3$ million citizens. Figure \ref{fig:ranking} shows this correlation ($r=0.44$, $p<0.005$).

\subsection{Temporal}
\begin{figure}[h]
\centering
\includegraphics[width=1.0\columnwidth]{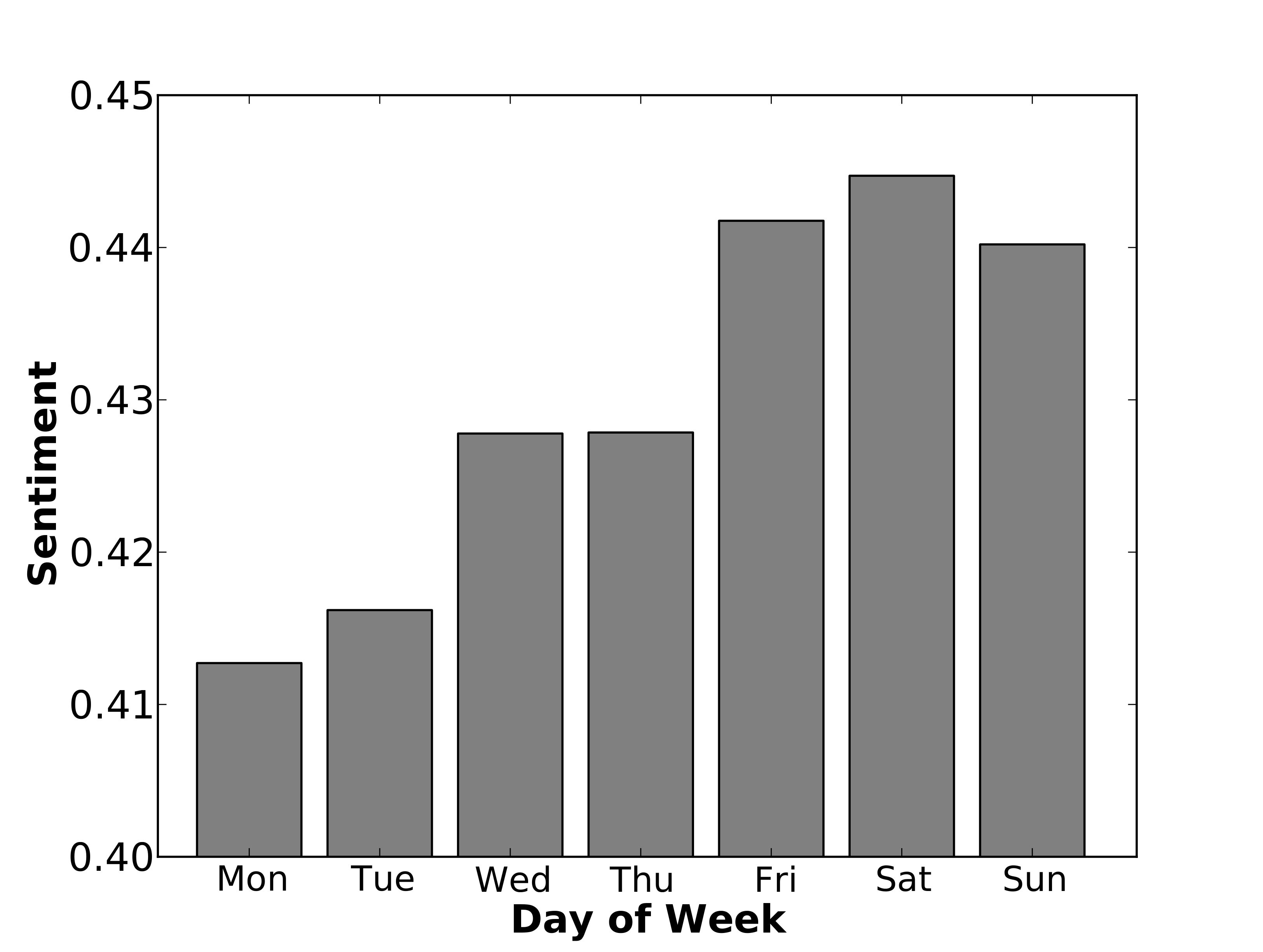}
\caption{Average sentiment of different days of the week in the USA, averaged across three years, from 2012 to 2014.}
\label{fig:dow}
\end{figure}

\begin{figure}[h]
\centering
\includegraphics[width=1.0\columnwidth]{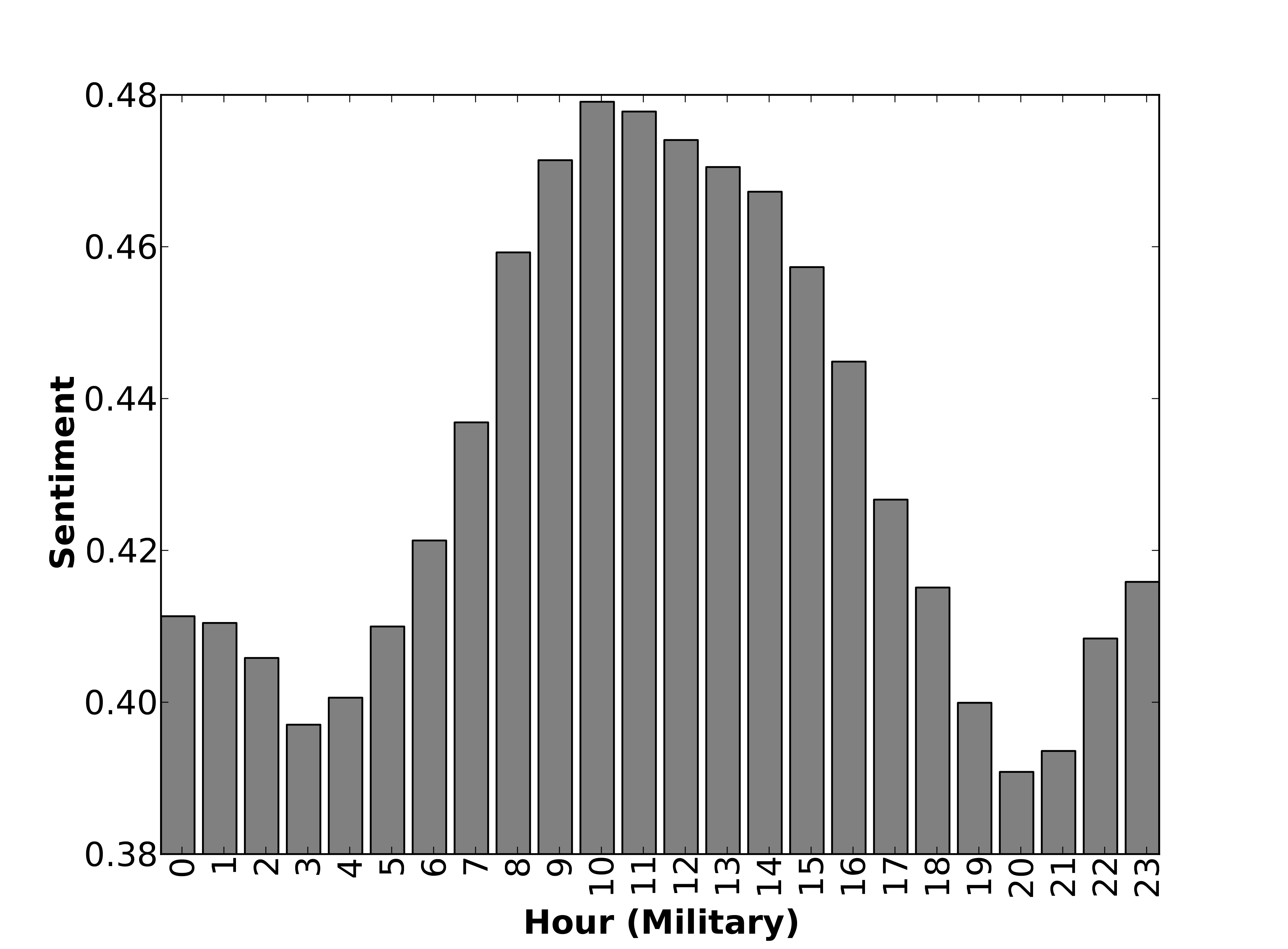}
\caption{Average sentiment of different hours of the day in the USA, averaged across three years, from 2012 to 2014.}
\label{fig:hour}
\end{figure}

\begin{figure}[h]
\centering
\includegraphics[width=1.0\columnwidth]{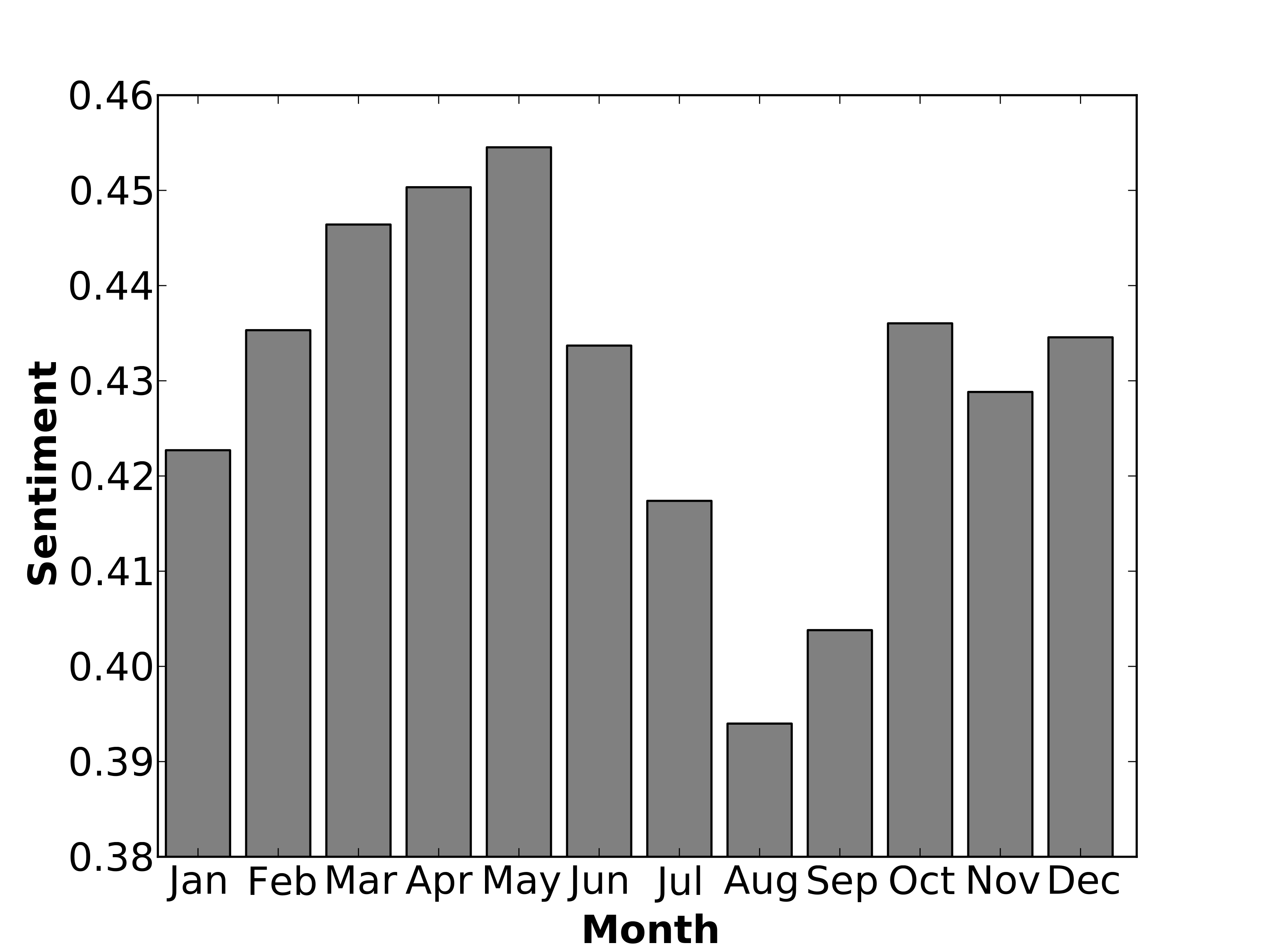}
\caption{Average sentiment of different months in the USA, averaged across three years, from 2012 to 2014.}
\label{fig:month}
\end{figure}

We looked at three temporal variables: time of day, day of the week and month. All tweets are tagged with timestamp data, which we used to extract these three variables. Since all timestamps in the Twitter historical archives (and public API) are in the UTC time zone, we first converted the timestamp to the local time of the location where the tweet was sent from. We then calculated the sentiment for each day of week (figure \ref{fig:dow}), hour (figure \ref{fig:hour}) and month (figure \ref{fig:month}), averaged across all $18$ million tweets over three years. The $18$ million tweets were divided evenly between each month, with $1.5$ million tweets per month. The tweets were also more or less evenly divided between each day of week, with each day having somewhere between $14\%$ and $15\%$ of the tweets. Similarly, the tweets were almost evenly divided between each hour, with each having somewhere between $3\%$ and $5\%$ of the tweets.

Some of these results make intuitive sense. For example, the closer the day of week is to Friday and Saturday, the more positive the sentiment, with a drop on Sunday. As with spatial, the average sentiment of all the hours, days and months lean more towards the positive side. 

\subsection{Authorial}
The last contextual variable we looked at was authorial. People have different baseline attitudes, some are optimistic and positive, some are pessimistic and negative, and some are in between. This difference in personalities can manifest itself in the sentiment of tweets. We attempted to capture this difference by looking at the history of tweets made by users. The $18$ million labelled tweets in our dataset come from $7,657,158$ authors. 

\begin{figure}[h]
\centering
\includegraphics[width=0.9\columnwidth]{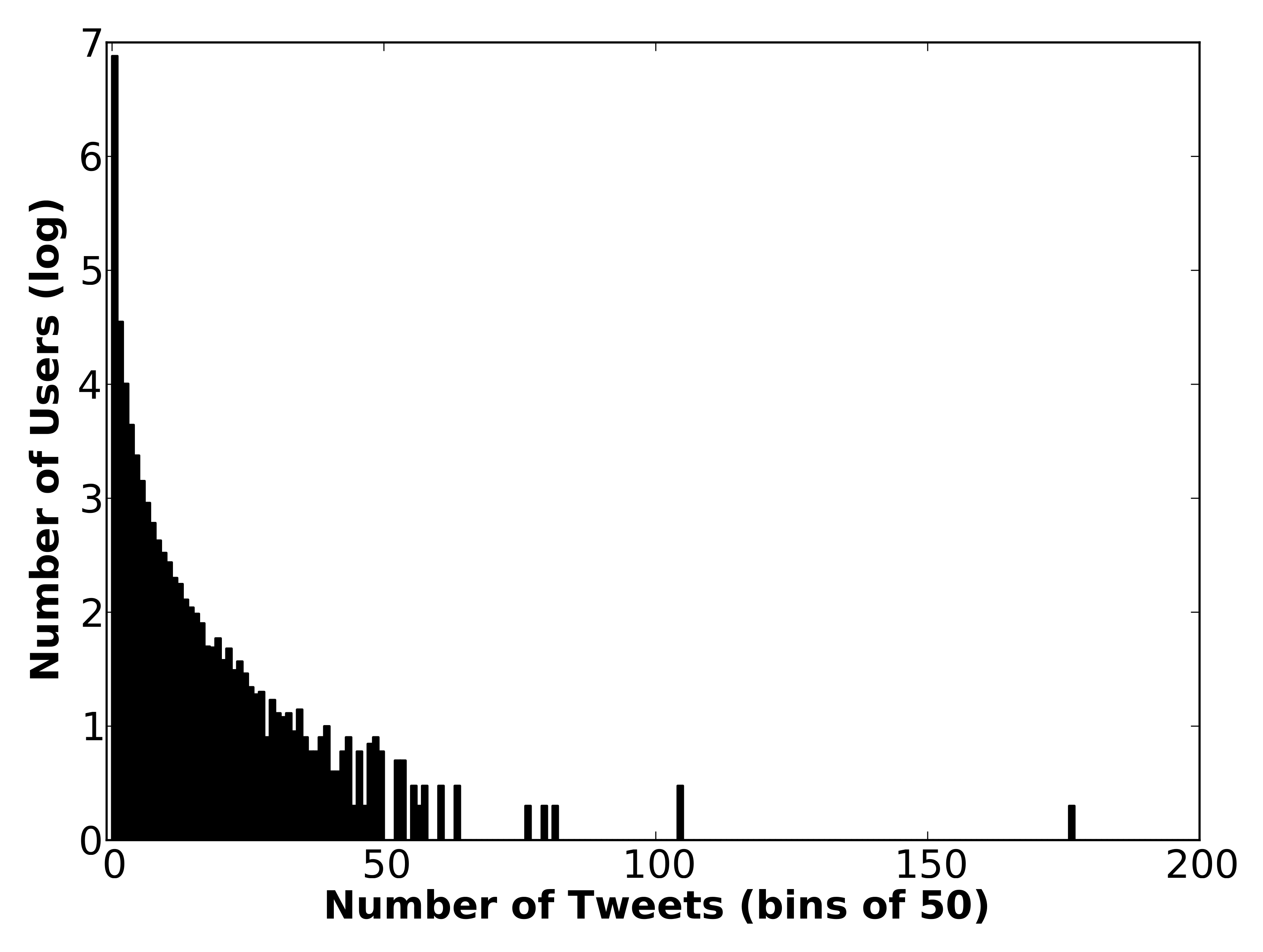}
\caption{Number of users (logarithmic) in bins of 50 tweets. The first bin corresponds to number of users that have less than 50 tweets throughout the three years and so on.}
\label{fig:author}
\end{figure}

In order to calculate a statistically significant average sentiment for each author, we need our sample size to not be too small. However, a large number of the users in our dataset only tweeted once or twice during the three years. Figure \ref{fig:author} shows the number of users in bins of 50 tweets. (So the first bin corresponds to the number of users that have less than 50 tweets throughout the three year.) The number of users in the first few bins were so large that the graph needed to be logarithmic in order to be legible. We decided to calculate the prior sentiment for users with at least $50$ tweets. This corresponded to less than $1\%$ of the users ($57,710$ out of $7,657,158$ total users). Note that these users are the most prolific authors in our dataset, as they account for $39\%$ of all tweets in our dataset. The users with less than $50$ posts had their prior set to $0.0$, not favouring positive or negative sentiment (this way it does not have an impact on the Bayesian model, allowing other contextual variables to set the prior).

\begin{figure}[h]
\centering
\includegraphics[width=0.9\columnwidth]{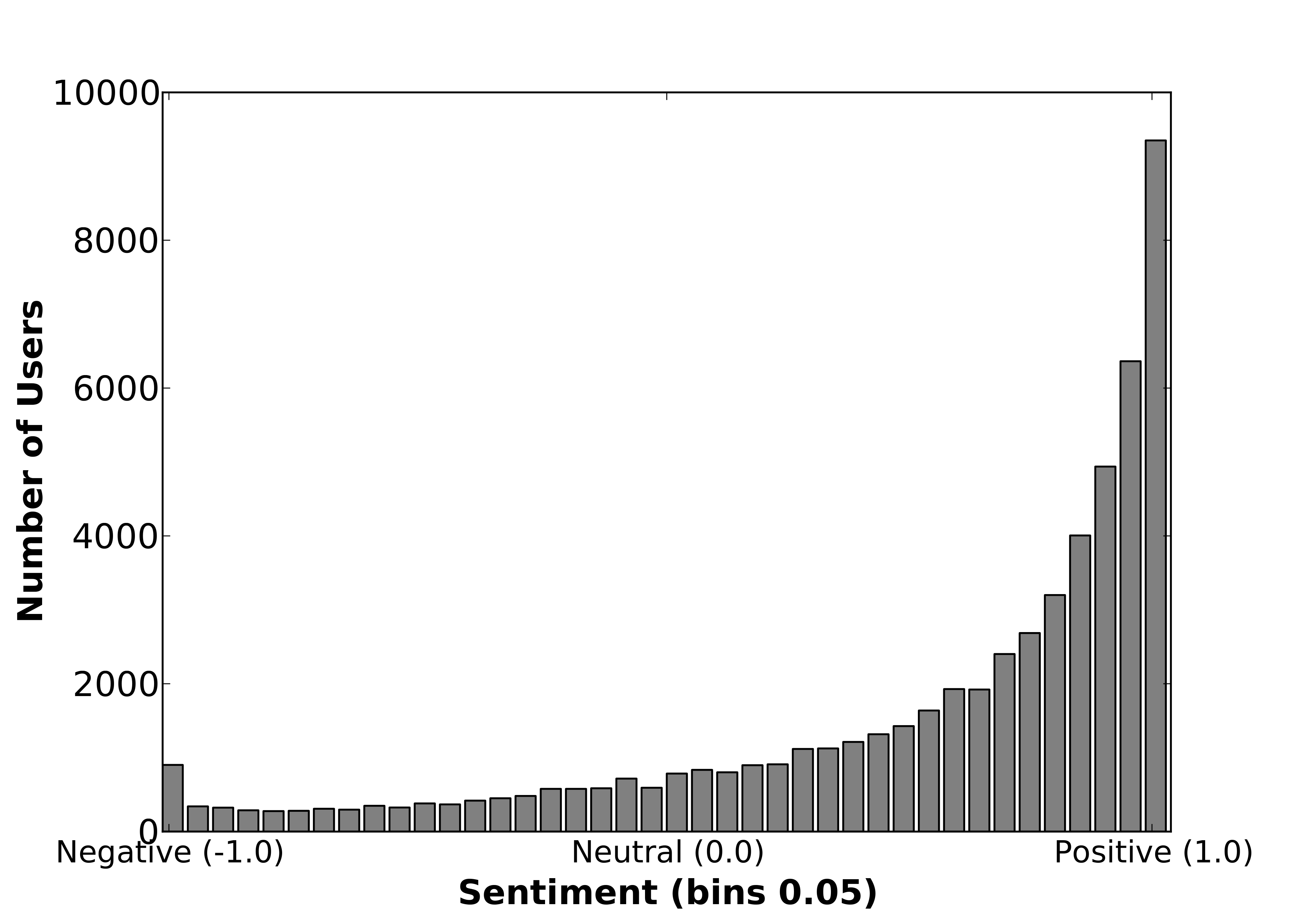}
\caption{Number of users (with at least 50 tweets) per sentiment bins of $0.05$, averaged across three years, from 2012 to 2014.}
\label{fig:author_sent}
\end{figure}

As it is not feasible to show the prior average sentiment of all $57,710$ users, we created $20$ even sentiment bins, from $-1.0$ to $1.0$. We then plotted the number of users whose average sentiment falls into these bins (Figure \ref{fig:author_sent}). Similar to other variables, the positive end of the graph is much heavier than the negative end.


\section{Results}
We used 5-fold cross validation to train and evaluate our baseline and contextual models, ensuring that the tweets in the training folds were not used in the calculation of any of the priors or in the training of the bigram models. Table \ref{tab:results} shows the accuracy of our models. The contextual model outperformed the baseline model using any of the contextual variables by themselves, with state being the best performing and day of week the worst. The model that utilized all the contextual variables saw a $10\%$ relative and $8\%$ absolute improvement over the baseline bigram model.

\begin{table}[h]
\centering
\begin{tabular}{l|l}
Model & Accuracy \\ \hline
Baseline-Majority & $0.620$  \\ \hline
Baseline-Bigram & $0.785$   \\ \hline
Contextual-DoW & $0.798$  \\ \hline
Contextual-Month & $0.801$  \\ \hline
Contextual-Hour & $0.821$  \\ \hline
Contextual-Author & $0.829$  \\ \hline
Contextual-State & $0.849$  \\ \hline
Contextual-All & $\mathbf{0.862}$   


\end{tabular}
\caption{Classifier accuracy, sorted from worst to best. }
\label{tab:results}
\end{table}

Because of the great increase in the volume of data, distant supervised sentiment classifiers for Twitter tend to generally outperform more standard classifiers using human-labelled datasets. Therefore, it makes sense to compare the performance of our classifier to other distant supervised classifiers. Though not directly comparable, our contextual classifier outperforms the distant supervised Twitter sentiment classifier by Go et al \cite{go2009twitter} by more than $3\%$ (absolute).

Table \ref{tab:results2} shows the precision, recall and F1 score of the positive and negative class for the full contextual classifier (Contextual-All). 

\begin{table}[h]
\centering
\begin{tabular}{l|l|l|l}
Class & Precision & Recall & F1 Score \\ \hline
Positive & $0.864$ & $0.969$  & $0.912$  \\ \hline
Negative  & $0.905$ & $0.795$ & $0.841$  \\ 

\end{tabular}
\caption{Precision, recall and F1 score of the full contextual classifier (Contexual-All).}
\label{tab:results2}
\end{table}

\section{Discussions} 


Even though our contextual classifier was able to outperform the previous state-of-the-art, distant supervised sentiment classifier, it should be noted that our contextual classifier's performance is boosted significantly by spatial information extracted through geo-tags. However, only about one to two percent of tweets in the wild are geo-tagged. Therefore, we trained and evaluated our contextual model using all the variables except for state. The accuracy of this model was $0.843$, which is still significantly better than the performance of the purely linguistic classifier. Fortunately, all tweets are tagged with timestamps and author information, so all the other four contextual variables used in our model can be used for classifying the sentiment of any tweet.

Note that the prior probabilities that we calculated need to be recalculated and updated every once in a while to account for changes in the world. For example, a state might become more affluent, causing its citizens to become on average happier. This change could potentially have an effect on the average sentiment expressed by the citizens of that state on Twitter, which would make our priors obsolete. 



\section{Conclusions and Future Work}

Sentiment classification of tweets is an important area of research. Through classification and analysis of sentiments on Twitter, one can get an understanding of people's attitudes about particular topics. 

In this work, we utilized the power of distant supervision to collect millions of noisy labelled tweets from all over the USA, across three years. We used this dataset to create prior probabilities for the average sentiment of tweets in different spatial, temporal and authorial contexts. We then used a Bayesian approach to combine these priors with standard bigram language models. The resulting combined model was able to achieve an accuracy of $0.862$, outperforming the previous state-of-the-art distant supervised Twitter sentiment classifier by more than $3\%$.


In the future, we would like to explore additional contextual features that could be predictive of sentiment on Twitter. Specifically, we would like to incorporate the topic type of tweets into our model. The topic type characterizes the nature of the topics discussed in tweets (e.g., breaking news, sports, etc). There has already been extensive work done on topic categorization schemes for Twitter \cite{dann2010twitter,sriram2010short,zhao2011empirical} which we can utilize for this task. 



\section{Acknowledgements}
We would like to thank all the annotators for their efforts. We would also like to thank Brandon Roy for sharing his insights on Bayesian modelling. This work was supported by a generous grant from Twitter.


\bibliographystyle{acl}
\bibliography{sv_emnlp}

\end{document}